\documentstyle[twoside, epsfig]{article}

\input ibvs2.sty

\begin{document}
\IBVShead{5xxx}{16 July 2015}

\IBVStitle{AO Psc time keeping}

\IBVSauth{Michel Bonnardeau$^1$}

\IBVSinst{MBCAA Observatory, Le Pavillon, 38930 Lalley, France, email: arzelier1@free.fr}

\SIMBADobjAlias{AO Psc}
\IBVStyp{GCVS}
\IBVSkey{photometry}
\IBVSabs{Eleven seasons, from 2004 to 2014, of photometric monitoring of the intermediate polar AO Psc are presented and are compared with previous observations. The spin up of the white dwarf is found to be slowing down. The amplitudes of the modulated and non-modulated components of the brightness are found to have undergone a major change in 2007. }

\begintext

AO Piscium (RA=22h 55min 17.99s DEC=-03° 10' 40.0" J2000.) is an intermediate polar, that is a subclass of cataclysmic systems in which the white dwarf is magnetized enough to module the accretion. Furthermore, the period of rotation (or spin) of the white dwarf is shorter than the orbital period and there is an accretion disc. AO Psc is one of the brightest cataclysmic, with a V mag as high as 13.2.

The orbital period is $P_{orb}=3.59$ $hr$, the rotation period of the white dwarf is $P_{rot}=805$ $s$ and the accretion X-ray beam is reprocessed on the secondary star atmosphere, giving rise to a synodic modulation with the period $P_{syn}$ such that:
\\
$1/P_{syn}=1/P_{orb}-1/P_{rot}$
\\
i.e. $P_{syn}=859$ $s$ 
(Patterson \& Price, 1981, Motch \& Pakull, 1981, van Amerongen et al., 1985 (hereafter vA85), Taylor et al, 1997). 

All these periodicities are visible by photometry as modulations in the light curves, the synodic modulation being usually the strongest one.\\ 

\IBVSfig{5cm}{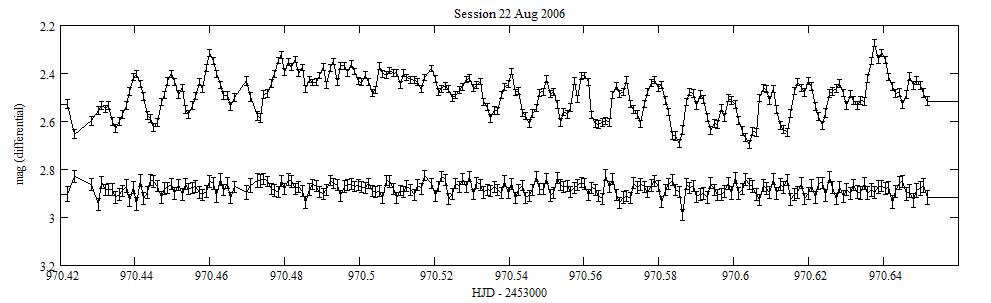}{Upper light curve: AO Psc, Lower: the check star shifted by -0.2 mag. The error bars are the quadratic sum of the 1-sigma statistical uncertainties on the variable/check star and on the comparison star.}
Photometric observations of AO Psc were carried out over eleven seasons, from 2004 to 2014, with a 203 mm f/6.3 Schmidt-Cassegrain telescope, a Clear filter and a SBIG ST7E camera (KAF401E CCD). The exposures were 60 s long. The images were dark substracted (using master darks of the same duration than the images and at the same temperatures) and flat corrected (MaximDL software program). For the aperture differential photometry (AstroMB software package), the comparison star is GSC 5238-462. A check star, GSC 5238-347, is used to compare the standard deviations to the statistical uncertainties so as to make sure that the systematic errors are low. An example of a light curve is given Figure 1. A total of 8744 images were obtained over 74 nights.\\

\vskip 0.5cm
\centerline{Table 1: Results of the fits and cycle counts}
\begin{center}
\begin{tabular}{|l|r|r|r|r|r|r|r|r|r|r|}
\hline {\small Season}  & $t_{syn}$ & $N_{syn}$ & $t_{rot}$ & $N_{rot}$ & $t_{orb}$ & $N_{orb}$ & $A_{0}$
& $A_{syn}$ & $A_{rot}$ & $A_{orb}$\\
\hline 2004 & {\small 322.3748} & {\small (a)} & {\small 323.2710} & {\small  (b)} & {\small 346.3534} & {\small  (c)} & {\small 2.614} &{\small  -0.120} & {\small -0.054} & {\small 0.033}\\
& {\small $\pm 10$} &  & {\small $\pm 22$} &   & {\small $\pm 68$} & 
& {\small $\pm 0.002$} & {\small $\pm 0.001$} & {\small $\pm 0.001$} & {\small $\pm 0.002$}\\
\hline 2005 & {\small 612.6533} & {\small 29,209} & {\small 612.6264} & {\small 31,050} & {\small 701.2596} & {\small 2,372} & {\small 2.488} &{\small  -0.117} & {\small -0.051} & {\small 0.063}\\
& {\small $\pm 7$} &  &    {\small $\pm 25$} & & {\small $\pm 40$} &
& {\small $\pm 0.001$} & {\small $\pm 0.001$} & {\small $\pm 0.001$} & {\small $\pm 0.001$}\\
\hline 2006 & {\small 970.4400} & {\small 36,002} & {\small 970.4118} & {\small 38,393} & {\small 970.5857} & {\small 1,800} & {\small 2.530} &{\small  -0.117} & {\small -0.063} & {\small 0.068}\\
& {\small $\pm 12$} &  &    {\small $\pm 13$} & & {\small $\pm 49$} &
& {\small $\pm 0.001$} & {\small $\pm 0.001$} & {\small $\pm 0.002$} & {\small $\pm 0.001$}\\
\hline 2007 & {\small 1301.5420} & {\small 33,317} & {\small 1356.4425} & {\small 41,424} & {\small 1296.6073} & {\small 2,179} & {\small 2.214} &{\small  -0.044} & {\small -0.016} & {\small 0.091}\\
& {\small $\pm 9$} &  &    {\small $\pm 36$} & & {\small $\pm 39$} &
& {\small $\pm 0.001$} & {\small $\pm 0.001$} & {\small $\pm 0.001$} & {\small $\pm 0.001$}\\
\hline 2008 & {\small 1709.5149} & {\small 41,052} & {\small 1681.5259} & {\small 34,884} & {\small 1681.6001} & {\small 2,573} & {\small 2.271} &{\small  -0.036} & {\small -0.038} & {\small 0.062}\\
& {\small $\pm 18$} &  &    {\small $\pm 15$} & & {\small $\pm 9$} &
& {\small $\pm 0.001$} & {\small $\pm 0.001$} & {\small $\pm 0.001$} & {\small $\pm 0.001$}\\
\hline 2009 & {\small 2041.5198} & {\small 33,408} & {\small 2041.5171} & {\small 38,630} & {\small 2041.5900} & {\small 2,406} & {\small 2.227} &{\small  -0.029} & {\small -0.028} & {\small 0.080}\\
& {\small $\pm 16$} &  &    {\small $\pm 21$} & & {\small $\pm 44$} &
& {\small $\pm 0.001$} & {\small $\pm 0.001$} & {\small $\pm 0.001$} & {\small $\pm 0.001$}\\
\hline 2010 & {\small 2415.5229} & {\small 37,634} & {\small 2454.5989} & {\small 44,327} & {\small 2415.5044} & {\small 2,499} & {\small 2.292} &{\small  -0.032} & {\small -0.040} & {\small 0.075}\\
& {\small $\pm 30$} &  &    {\small $\pm 22$} & & {\small $\pm 33$} &
& {\small $\pm 0.002$} & {\small $\pm 0.002$} & {\small $\pm 0.001$} & {\small $\pm 0.001$}\\
\hline 2011 & {\small 2744.5262} & {\small 33,106} & {\small 2748.5909} & {\small 31,548} & {\small 2744.5315} & {\small 2,199} & {\small 2.207} &{\small  -0.041} & {\small -0.024} & {\small 0.100}\\
& {\small $\pm 17$} &  &    {\small $\pm 16$} & & {\small $\pm 23$} &
& {\small $\pm 0.001$} & {\small $\pm 0.001$} & {\small $\pm 0.001$} & {\small $\pm 0.001$}\\
\hline 2012 & {\small 3140.4698} & {\small 39,842} & {\small 3140.4705} & {\small 42,052} & {\small 3126.5269} & {\small 2,553} & {\small 2.218} &{\small  -0.028} & {\small -0.020} & {\small 0.063}\\
& {\small $\pm 38$} &  &    {\small $\pm 35$} & & {\small $\pm 13$} &
& {\small $\pm 0.001$} & {\small $\pm 0.001$} & {\small $\pm 0.001$} & {\small $\pm 0.001$}\\
\hline 2013 & {\small 3489.5265} & {\small 35,124} & {\small 3489.4897} & {\small 37,453} & {\small 3559.3922} & {\small 2,893} & {\small 2.189} &{\small  -0.049} & {\small -0.017} & {\small 0.083}\\
& {\small $\pm 10$} &  &    {\small $\pm 51$} & & {\small $\pm 61$} &
& {\small $\pm 0.001$} & {\small $\pm 0.001$} & {\small $\pm 0.001$} & {\small $\pm 0.001$}\\
\hline 2014 & {\small 3836.4857} & {\small 34,913} & {\small 3836.4863} & {\small 37,236} & {\small 3865.5218} & {\small 2,046} & {\small 2.173} &{\small  -0.067} & {\small -0.007} & {\small 0.087}\\
& {\small $\pm 11$} &  &    {\small $\pm 72$} & & {\small $\pm 72$} &
& {\small $\pm 0.001$} & {\small $\pm 0.001$} & {\small $\pm 0.003$} & {\small $\pm 0.001$}\\
\hline
\end{tabular}
\end{center}
\leftline{the $t_{xxx}$ are in HJD - 2,453,000 with the uncertainties in seconds,}\leftline{the $N_{xxx}$ for one season is the difference with the previous season,}\leftline{the $A_{xxx}$ are in mag.}
(a) 819,882 cycles from the 0 of vA85, 83,170 cycles from the 2002 measurement of Williams (2003) (hereafter W03).\\
(b) 905,581 cycles from the 0 of vA85, 711,933 cycles from the 1986 measurement of Kaluzny \& Semeniuk (1988) (hereafter KS88).\\
(c) 56,689 cycles from the 0 of vA85 + $P_{orb}/2$, 44,495 cycles from the 1986 measurement of KS88 + $P_{orb}/2$. \\


The magnitudes as a function of time t are fitted by the following H(t) function: 
\\
$H(t) = A_{0} + H_{syn}(t) + H_{rot}(t) + H_{orb}(t) $
\\
where $A_{0}$ is a constant, $H_{syn}(t)$ is the synodic modulation: 
\\
$H_{syn}(t) = A_{syn}[cos(\pi(t - t_{syn})/P_{syn}]^2$ 
\\
$H_{rot}(t)$ is the rotation modulation: 
\\
$H_{rot}(t) = A_{rot}[cos(\pi(t - t_{rot})/P_{rot}]^2$ 
\\
and $H_{orb}(t)$ is the orbital modulation: 
\\
$H_{orb}(t) = A_{orb}[1+cos(2\pi(t-t_{orb})/P_{orb})]$ 

The $H(t$) function is fitted to the observations owing to a Monte Carlo method to test the parameters relative to the timing and, for each trial, the amplitudes are determined by a least squares method. The magnitudes are weighted with the uncertainties.

Each Monte Carlo run is made of 10 millions trials. The averages and standard deviations for 10 runs are given in Table 1, along with the number of cycles,  Nxxx. 

In 2007 the synodic and rotation modulations become fainter, especially the synodic one, and the orbital modulation and the non-modulated part $A_{0}$ become brighter, as shown Figures 2-5.

\IBVS2fig{6.0cm}{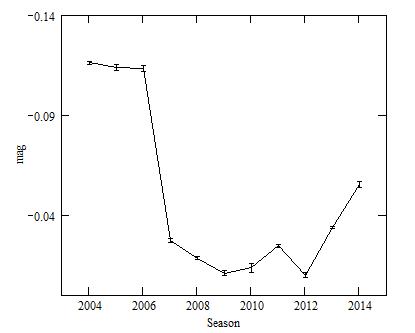}{The amplitude $A_{syn}$}{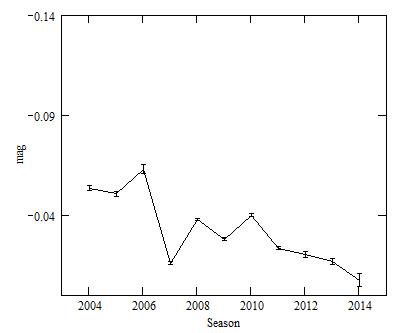}{The amplitude $A_{rot}$}
\IBVS2fig{6.0cm}{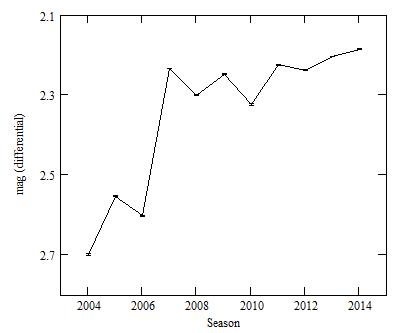}{The amplitude $A_{0}$}{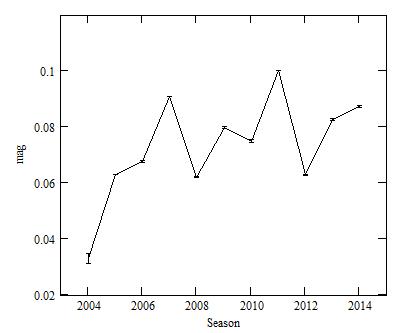}{The amplitude $A_{orb}$}

The times of maxima of the synodic modulation may be fitted with the function $ToM(n)=T_{syn}+P_{syn}n+b_{syn}n^2$. There are 26 such maxima (9 from vA85, 1 from KS88, 5 from W03 and 11 from this paper). With only the data from vA85 and of W03, there is an ambiguity in the cycle count and two fits are possible (W03). Indeed, the residuals (weighted with the uncertainties) are then 8.1 s for the Fit 1 of W03 and 10.4 s for the Fit 2. But adding the data presented here allow lifting the ambiguity: 8.9 s for the Fit 1, 15.6 s for the Fit 2. Adding the measurement of KS88 gives 9.2 s and 15.4 s respectively. Therefore, the cycle count of the Fit 1 of W03 is the right one.

The fit of the 2004-14 synodic maxima gives $b_{syn}=-(2.544 \pm 0.043).10^{-13}$ day. This is larger (smaller in absolute value) that the fits of W03, themselves larger than the ones of KS88 and of vA85. All the 26 maxima are then fitted with a supplementary term, $\gamma_{syn}n^3$. Furthermore, they are corrected for the leap seconds due to the Earth rotation slowing down (Eastman et al, 2010). For $T_{syn}$ in 1982 this correction is 21 s, for the first maximum the correction is 19 s, 35 s for the last one. $T_{syn}$ is to be expressed in HJD, the corrections are then between -2 s and +14 s. The barycentric effect of Jupiter and Saturn is neglected as it is only $\pm$ 4 s and cyclic (unlike the leap seconds that keep accumulating), and the other general relativistic corrections are much smaller. The least squares method gives: \\
$b_{syn} = -3.020.10^{-13}$ $day$  \\
$\gamma_{syn} = 1.44.10^{-21}$ $day$. 

The fit is also done with a Monte Carlo method, so as to have a result that is independent from the least squares method and to evaluate the uncertainties. For a Monte Carlo run, $T_{syn}$, $P_{syn}$, $b_{syn}$ and $\gamma_{syn}$ are chosen each in its own range; for $\gamma_{syn}$ the range is $[-10,+10].10^{-21}$. 10 millions trials are computed for a run. The averages and standard deviations of 10 runs are: \\
$T_{syn} = 2,445,174.181,13(2)$ $HJD$ \\
$P_{syn} = 0.009,938,498,0(4)$ $day$ \\
$b_{syn} = -3.031(8).10^{-13}$ $day$ \\
$\gamma_{syn} = 2.13(44).10^{-21}$ $day$.

Therefore the spinning up is slowing down. The derivative of the period is:\\
$\dot{P}_{syn} = 2b_{syn}/P_{syn} = -6.10.10^{-11}$ \\ 
and the secondary derivative of the period is:\\
$\ddot{P}_{syn} = 6\gamma_{syn}/P_{syn}^2 = 1.30.10^{-16}$ $day^{-1}.$\\
This gives the time scale:\\
$\tau = P_{syn}/(2\dot{P}_{syn}) = -223$ $kyr$ \\
and the breaking index:\\
$n = P_{syn}\ddot{P}_{syn}/\dot{P}_{syn}^2 = 346$.\\
(By comparison, for FO Aqr, one has from W03: $\tau = 194$ $kyr$, $n = -6431$).
\\

There are 7 orbital maxima from vA85 and one from KS88. In order to fit them with the 11 orbital minima presented here, they are corrected by adding $P_{orb}/2$. A Monte Carlo method (rather than a least squares method, so as to evaluate the uncertainties) gives the ephemeris, for the orbital minima, taking into account the leap seconds:\\ 
$t(n) = T_{orb}+P_{orb}n $ \\
with:\\
$T_{orb} = 2,444,864.21809(1)$ $HJD$ \\
$P_{orb} = 0.149,625,502,2(1)$ $day$ \\
This is within the error bars of the ephemeris of KS88, with a better precision. An ephemeris with a quadratic term was also searched for, but with no significant improvement. 
\\

As the rotation modulation is related to the synodic modulation and the orbital period, the number of rotations of the white dwarf may be calculated unambiguously. The results are given in Table 1 at (b).  

\references
Eastman J., Siverd R. and Gaudi B.S., 2010, \textit{PASP} \textbf{122} 935.

Kaluzny J. and Semeniuk I., 1988, \textit{IBVS} 3145. 

Motch C. and Pakull M.W., 1981, \textit{A\&A} \textbf{101} L9. 

Patterson J. and Price C.M., 1981, \textit{ApJ} \textbf{243} L83.

Taylor P., Beardmore A.P., Norton A.J., Osborne J.P. and Watson M.G., 1997, \textit{MNRAS} \textbf{289} 349.

van Amerongen S., Kraakman H., Damen E., Tjemkes S. and van Paradijs J., 1985, \textit{MNRAS} \textbf{215} 45. 

Williams G., 2003, \textit{PASP} \textbf{115} 618. 

\endreferences

\end{document}